\documentclass[useAMS,usenatbib]{mn2e}

\usepackage{graphicx,epsfig,subfigure}
\usepackage{url}

\usepackage{amsmath}
\usepackage{amssymb}

\usepackage{threeparttable}
\usepackage{float}
\usepackage{morefloats}

\title[Dark Matter in ETGs]
   {Dark matter inside early-type galaxies as function of mass and redshift.}

   \author[Nigoche-Netro et al.]
  {A. Nigoche-Netro,$^1$\thanks{E-mail: anigoche@gmail.com}
  G. Ramos-Larios,$^1$ P. Lagos,$^2$
  \newauthor
  A. Ruelas-Mayorga,$^3$
  E. de la Fuente,$^1$ S. N. Kemp,$^1$
  \newauthor
  S. G. Navarro,$^1$
  L. J. Corral,$^1$ A. M. Hidalgo-G\'amez, $^4$ \\
    $^1$Instituto de Astronom\'{\i}a y Meteorolog\'{\i}a, Universidad de Guadalajara, Guadalajara, Jal. 44130, M\'exico.\\
    $^2$Instituto de Astrof\'isica e Ci\^encias do Espa\c{c}o, Universidade do Porto, Rua das Estrelas, 4150-762 Porto, Portugal.\\
    $^3$Instituto de Astronom\'{\i}a, Universidad Nacional Aut\'onoma de M\'exico, Cd. Universitaria, M\'exico, D.F. 04510, M\'exico.\\
    $^4$Escuela Superior de F\'{\i}sica y Matem\'aticas, Instituto Polit\'ecnico Nacional, M\'exico, D. F., M\'exico.\\}

\begin{document}

\date{Accepted 2015. Received 2015 October; in original form 2015 October}

\pagerange{\pageref{firstpage}--\pageref{lastpage}}
\pubyear{2012}

\maketitle

\label{firstpage}

\begin{abstract}

We study the behaviour of the dynamical and stellar mass inside the effective radius ($r_{e}$) of early-type galaxies (ETGs). We use several samples of ETGs -ranging from 19 000 to 98 000 objects- from the ninth data release of the Sloan Digital Sky Survey.  We consider Newtonian dynamics, different light profiles and different Initial Mass Functions (IMF) to calculate the dynamical and stellar mass. We assume that any difference between these two masses is due to dark matter and/or a non Universal IMF. The main results for galaxies in the redshift range $0.0024 < z < 0.3500$ and in the dynamical mass range 9.5 $<$ log(M) $<$ 12.5 are: i) A significant part of the intrinsic dispersion of the distribution of dynamical vs. stellar mass is due to redshift. ii) The difference between dynamical and stellar mass increases as a function of dynamical mass and decreases as a function of redshift. iii) The difference between dynamical and stellar mass goes from approximately 0\% to 70\% of the dynamical mass depending on mass and redshift. iv) These differences could be due to dark matter or a non Universal IMF or a combination of both. v) The amount of dark matter inside ETGs would be equal to or less than the difference between dynamical and stellar mass depending on the impact of the IMF on the stellar mass estimation. vi) The previous results go in the same direction of some results of the Fundamental Plane (FP) found in the literature in the sense that they could be interpreted as an increase of dark matter along the FP and a dependence of the FP on redshift.

\end{abstract}




\section{Introduction}

\label{sec:intro}

The measurement of masses of galaxies has been, over a long period of time, an interesting and difficult problem, which has elicited the application of various and diverse techniques \citep{spi75,bur75,sof01,sim07}. Since the determination of rotation curves for a large number of spiral galaxies \citep{sof01} and the suggestion that these rotation curves are flat because of the presence of an unseen amount of mass which has been called `dark matter', the determination of the mass of all types of galaxies has become a pressing concern of modern Astronomy. It is fair to say at this point that there is no direct evidence of the existence of dark matter and that there are other explanations which, although not as currently popular as dark matter, may explain the observations quite reasonably.

The total mass of a galaxy is composed of two elements; luminous matter and dark matter. If we assume that both luminous and dark matter respond to the Newtonian gravitational law in the same way, then the difference between the dynamical mass and the luminous mass of a galaxy provides us with an estimation of the amount of dark matter present in the galactic system in question. From such a determination we would be able to study if a dependence of the amount of dark matter with dynamical mass and/or redshift exists.

Measuring the amount of radiation from a particular galaxy, combined with typical mass to light ratios (${\bf M}$/$L$) that have been calibrated using different stellar samples in our own Galaxy, allows us to estimate its stellar, gas and dust content. Moreover, rotation curves for spiral galaxies permit the calculation of dynamical mass inside any radius for which a value of rotation velocity is known, allowing us, in principle, to calculate from these two determinations the amount of dark matter present in the galaxy under study. As is well known, rotation velocity curves are used for studying the kinematics of galaxies, determining the amount and distribution of mass interior to a given radius,  to derive an insight into galactic evolutionary histories and the possible role that interactions with other systems may have played. Since rotation curves may be obtained at different wavelengths they provide information as to the kinematics of different constituents of a galaxy. They may be observed in the infrared as well as in the optical, which may be used to trace ionised gas and the stellar motions, also in the radio and microwave regimes which trace the neutral and molecular gas components of a galaxy.

Recently, stellar population synthesis models have been used to calculate galactic masses. These models also give us an idea of the total stellar content of a galaxy as well as the distribution of stars of all the different spectral types and luminosity classes \citep{aug10,bar11,son12}.

Dynamical theoretical models can also be used to calculate masses for early-type galaxies (ETGs), such as those which \citet{vandermarel1991} constructed for 37 bright elliptical galaxies. From these models he found an average (M/$L)_B=(5.93 \pm 0.25)h_{50}$. Discrepancies of the observed velocities in the outer parts with those predicted by the models may be explained by the inclusion of massive dark haloes.

\citet{gerhard2001} performed dynamical studies of the shapes of line-profiles for 21 elliptical galaxies; they used them to investigate the dark halo properties and dynamical family relations of these galaxies. They appear to have minimal haloes implied from the fact that the ratio  M/$L_B$ turned out maximal. Some of these galaxies showed no dark matter within $2r_e$. \citet{cap06} investigated the correlations between the mass-to-light M/$L$ ratios of 25 elliptical and lenticular galaxies. Field and cluster galaxies presented no difference, and their dark matter content within an effective radius $r_e$ was $\sim 30\%$ of the total mass contained there. It appeared that the amount of dark matter correlates with galactic rotation velocity; in the sense that more massive slow-rotating galaxies contain less dark matter that the fast-rotating galaxies.

There have been many papers in which dynamical arguments are used to calculate the dynamical mass of galaxies, and hence, by comparison with the amount of luminous mass, they calculate the amount of dark matter present, see for example: \citet{thomas2007}, \citet{williams2009}, \citet{thomas2011}, \citet{cap13} to mention a few. Also check the detailed introduction to this topic published in \citet{nig15}.

The gravitational lens phenomenon provides direct and precise measurements of masses of galaxies at different scales, and allows us to establish the nature and presence of dark matter in a galactic system. Elliptical galaxies have been considered to have extended dark-matter massive haloes \citep{tre10} that follow the \citet{nav96} density profiles. \citet{ber93} and \citet{hum06} have studied the kinematics of different components in nearby elliptical systems and have concluded that dark matter haloes are required to explain the dynamics of massive elliptical galaxies, provided that Newtonian gravity be valid at these scales.

Galactic mass determinations have also been made using weak and strong lensing observations \citep{lag10,hoe05,gav07}.

The fraction of total mass in the form of dark matter in ETGs, $f_{DM}$, appears to increase with growing radius reaching values of $\sim 70\%$ at five effective radii \citep{tre04}. Furthermore, $f_{DM}$ within a fixed radius seems to grow with galaxy stellar mass and with velocity dispersion \citep{tor09,nap10,gra10,aug10a}. $f_{DM}$ varies from small values as in the case of bright giant elliptical galaxies \citep{rom03} to very large values, as has been found for dwarf spheroidal galaxies by \citet{sim07}. Studies of the Virgo giant elliptical galaxy NGC 4949 (M60) by \citet{teo11} reveal that the kinematics of Planetary Nebulae in this object is consistent with the presence of a dark matter halo with $f_{DM} \sim 0.5$ for $r=3r_e$. \citet{deb01} presented three-integral axisymmetric models for NGC 4649 and NGC 7097 and concluded that the kinematic data for NGC 4649 only require a small amount of dark matter, however \citet{das11} determine $f_{DM} \sim 0.78$ at $r=4r_e$ for NGC 4649.

Using gravitational lensing experiments, \citet{koopmans2006} find a projected dark matter fraction of $<f_{DM}>=0.25 \pm 0.06$ for 15 ETGs, while \citet{bar11} studying sixteen early-type lens galaxies determine the lower limit for dark matter $f_{DM}$ inside the effective radius. The median value for this fraction is $12 \%$ with variations from almost 0 to up to $50 \%$.

As mentioned above, direct detection of dark matter has not been achieved yet. Its presence requires the validity of Newtonian gravity. If we were to assume that at these very low acceleration regimes Newtonian gravity is not valid or may be slightly modified \citep{mil83} then further developments have explained several phenomena without the need of dark matter e.g. spiral galaxies, flat-rotation curves \citep{san02}, projected surface density profiles and observational parameters of the local dwarf spheroidal galaxies \citep{her10,mcg10,kro10}, the relative velocity of wide binaries in the solar neighbourhood \citep{her12}, fully self-consistent equilibrium models for NGC 4649 \citep{jim13} and references within among others.

In this paper we present a study of luminous and dynamical mass inside the effective radius of ETGs considering Newtonian dynamics. We search for differences between these masses and assume that any difference is due to dark matter or a non-Universal IMF or a combination of both.

The structure of this study is as follows; in \S 2 we present the sample of ETGs used in this work, in \S 3 we discuss the calculation of the stellar and virial masses for the galaxies in the sample, in \S 4 we discuss the distribution of stellar mass as a function of virial mass, in \S 5 and \S 6 we outline the difference between virial and stellar mass as a function of mass and redshift, in \S 7 we discuss our results in the Fundamental Plane context and finally in \S 8 we present the conclusions.

\section{The sample of ETGs}

We use a sample of ETGs from the Ninth Data Release (DR9) of the Sloan Digital Sky Survey (SDSS) \citep{yor00,aba09,aih11} and two subsamples of it, all of them in the $g$ and $r$ filters. These samples were compiled by and described in great detail in \citet{nig15}. Here we shall describe briefly the selection criteria used.

1) The brightness profile of the galaxy must be well adjusted by a de Vaucouleurs profile, in both the $g$ and $r$ filters (fracdevg = 1 and fracdevr = 1 according to the SDSS nomenclature).

2) The de Vaucouleurs magnitude of the galaxies must be contained in the interval $14.5 < m_{r, dev} < 17.5$ and its equivalent in the g filter.

3) The quotient of the semi axes (b/a) for the galaxies must be larger than 0.6 in both filters $g$ and $r$.

4) The galaxies must have a velocity dispersion of $\sigma_0 >$ 60 $km/s$ and a signal-to-noise ratio (S/N) $>$ 10.

The main sample is called ``Total-SDSS-Sample". It contains approximately 98000 galaxies, is distributed in a redshift interval $0.0024 < z < 0.3500$ and within a magnitude range $<\Delta M>$ $\sim 7$ $mag$ ($-17.5 \ge M_{g} > -24.5$). The first subsample is named ``The-Morphological-Sample". The main characteristic of The-Morphological-Sample is that the selection criteria for the morphological classification are more rigorous than in the Total-SDSS-Sample, due to the fact that \citet{nig15} use the morphological classification from the Galaxy Zoo project (see \citet{lin08}). With these added criteria they obtain approximately 27,000 ETGs. The last subsample is named ``The-Homogeneous-SDSS-Sample". In this case \citet{nig15} consider a volume limited sample (0.04 $\leq\;z\;\leq$ 0.08) with the objective of obtaining a complete sample in the bright end of the magnitude range. In this volume they obtain approximately 19 000 ETGs. This subsample covers a magnitude range $<\Delta M>$ $\sim 4.5$ $mag$ ($-18.5 \ge M_{g} > -23.0$) and is approximately complete for $M_{g} \leq -20.0$ (see Nigoche-Netro et al. 2015 for details).

The photometry and spectroscopy of the samples of galaxies were corrected due to different biases. Below we list these corrections:

\begin{itemize}

\item Seeing correction: The seeing-corrected parameters were obtained from the SDSS pipeline.

\item Extinction correction: The extinction correction values were obtained from the SDSS pipeline.

\item K correction: The K correction was obtained from \citet{nig08}.

\item Cosmological dimming correction: The cosmological dimming correction was obtained from \citet{jor95a}.

\item Evolution correction: The evolution correction was obtained from \citet{ber03b}.

\item Effective radius correction: The effective radius correction to the rest reference frame was obtained from \citet{hyd09}.

\item Aperture correction to the velocity dispersion: The velocity dispersion inside the radius subtended by the SDSS fibre was corrected using the  aperture correction from \citet{jor95b}.

\end{itemize}

 \section{The stellar and virial mass of the ETGs}

 We use the stellar and virial masses obtained in \citet{nig15}. Here we shall describe briefly the procedure used to calculate those masses and some terms that are important for the present work.

\subsection{The stellar mass}

The total stellar mass was obtained by \citet{nig15} considering different stellar population synthesis models, using a universal IMF (Salpeter or Kroupa) and different brightness profiles (de Vaucouleurs or S\'ersic). The combination of these ingredients results in three mass estimations, as follows:

i) de Vaucouleurs Salpeter-IMF stellar mass.

ii) S\'ersic Salpeter-IMF stellar mass.

iii) Kroupa-IMF stellar mass.

According to \citet{sch10}, within a sphere of radius equal to $r_{e}$, 42\% of the total stellar mass is contained.

The stellar masses described before assume a universal IMF. However, some papers in the astronomical literature claim that the IMF is not universal but rather it depends on the stellar mass \citep{cap12,dut13}. We do not correct the stellar mass for the behaviour of the IMF as a function of mass because there is no accurate equation describing this effect. Since our results have to take into account this effect, we will discussed them in the subsequent sections.

\subsection{The virial mass}

The total virial mass was obtained by \citet{nig15} using an equation from \citet{pov58}. This method assumes Newtonian mechanics and virial equilibrium for the galaxies in question. The equation is as follows:

\begin{equation}  \label{eq:eq2}
{\bf M_{virial}} \sim  K \frac{ r_{e} \sigma_{e}^{2}}{G},
\end{equation}

where the variables ${\bf M_{virial}}$, $r_{e}$ and, $\sigma_{e}$ represent respectively the total virial mass, the effective radius and the velocity dispersion inside $r_{e}$. $G$ stands for the gravitational constant and $K$ is a scale factor. For the de Vaucouleurs profile case $K=5.953$ \citep{cap06}. The amount of mass within an effective radius corresponds to 0.42 times the value calculated from equation (1). This mass may or may not be luminous.

The errors calculated for the different parameters reported in this paper are obtained using the rules of error propagation and considering possible systematics on the photometric and spectroscopic parameters as discussed in detail by \citet{nig15}.

In the following sections, and taking into consideration only the region internal to $r_{e}$, we will carry out an analysis of the behaviour of the virial vs. stellar mass.

 \section{Distribution of the stellar mass as a function of the virial mass of ETGs}

In \citet{nig15} we have made a complete analysis of the distribution of the stellar mass with respect to the virial mass for ETGs samples. In this section we present an extract of that analysis only with the relevant information for the goals of the present work. Figure 1 is the most important part of the extract because it shows the comparison of viral and stellar mass for each galaxy in our samples. In this figure, Column 1 represents the Total sample, Column 2 the Morphologic sample and Column 3 the Homogeneous sample. The rows correspond to different profiles and IMFs, being the first one associated with the de Vaucouleurs Salpeter-IMF stellar mass, the second with the Sersic Salpeter-IMF stellar mass and the third with the Kroupa-IMF stellar mass. The solid line is the one-to-one line.


\citet{nig15} discuss different procedures to analyse the distribution of masses shown in Figure 1 considering that the mentioned distribution may depend on observational biases, on physical properties of the galaxies, and on arbitrary cuts performed in the observed samples \citep[see also][]{nig08,nig09,nig10,nig11}. Those procedures may be helpful in investigating whether there is dark matter inside ETGs. Particularly they found that the application of the weighted bisector fit ($WBQ$ fit) to the mean value of the distribution at quasi-constant mass, results in a reduction of the possible biases which may creep in the process \citep[for details see section 7.3 of Appendix A from][]{nig15}. This method is only a first approximation to the study of dark matter inside ETGs because the distribution of masses seen in Figure 1 have a high intrinsic dispersion and the physical causes of the intrinsic dispersion are at present yet unknown. Refinements to this method must be sought for in the physical causes of the intrinsic dispersion seen in the mass distributions.


\begin{figure*}
   \begin{center}

      \includegraphics[width=17cm]{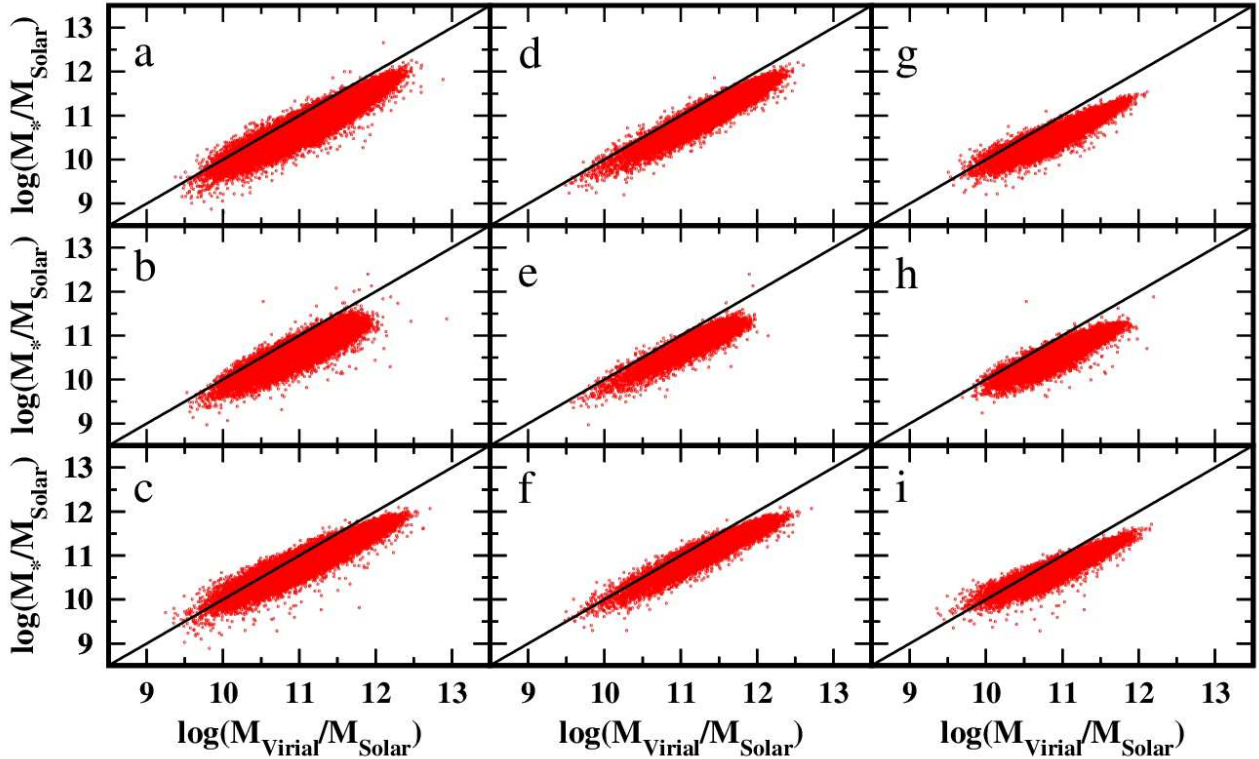}

         \caption{Distribution of the virial and stellar mass from the Total (column 1), Morphologic (column 2) and Homogeneous (column 3) samples. The first row corresponds to the de Vaucouleurs Salpeter-IMF stellar mass, the second one corresponds to the S\'ersic Salpeter-IMF stellar mass and the third one corresponds to the Kroupa-IMF stellar mass. The solid line is the one-to-one line.}

         \label{Fig1}
         \end{center}
   \end{figure*}


\begin{figure*}
   \begin{center}

      \includegraphics[width=17cm]{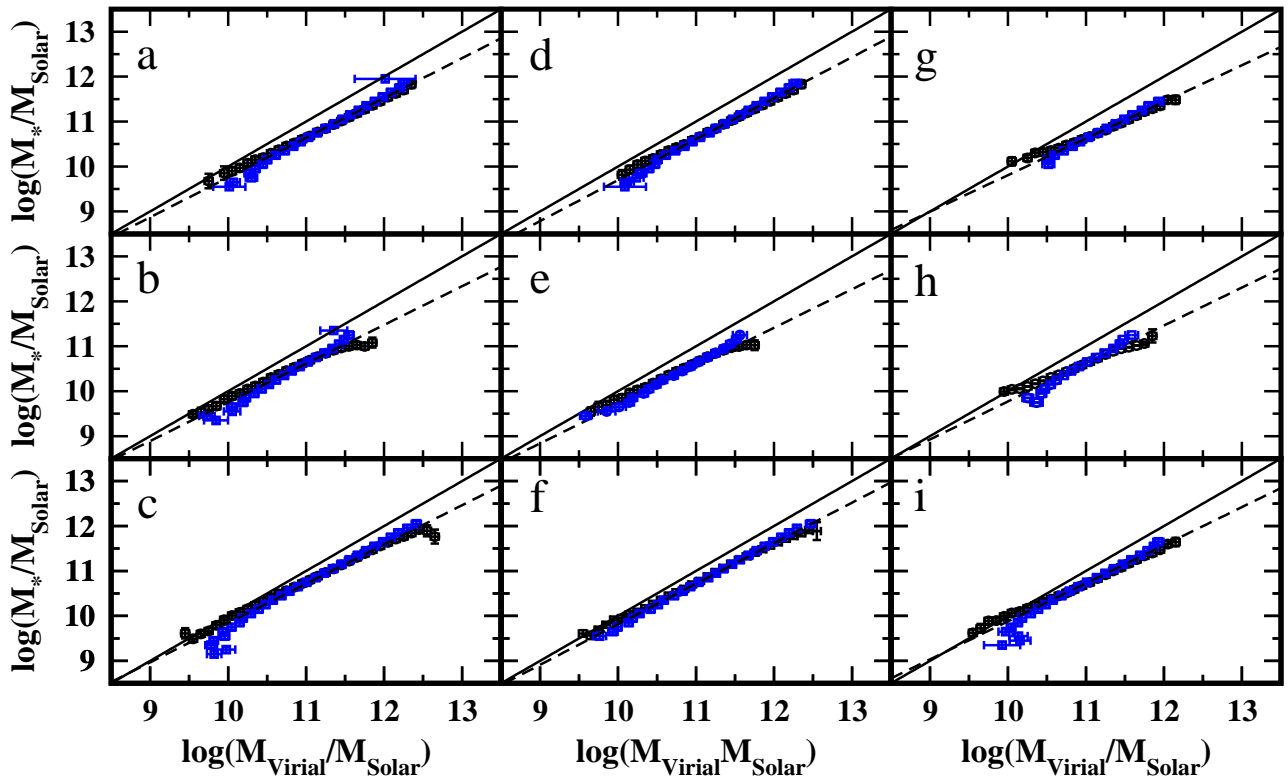}

         \caption{Distribution of the mean values of the virial and stellar mass from the Total (column 1), Morphologic (column 2) and Homogeneous (column 3) samples. The first row corresponds to the de Vaucouleurs Salpeter-IMF stellar mass, the second one corresponds to the S\'ersic Salpeter-IMF stellar mass and the third one corresponds to the Kroupa-IMF stellar mass. The black dots represent the mean values of the stellar mass at quasi-constant virial mass and the blue squares represent the mean values of the virial mass at quasi-constant stellar mass. The dashed line corresponds to the $WBQ$ fit. The solid line is the one-to-one line.}

         \label{Fig2}
         \end{center}
   \end{figure*}

\begin{table*}

\renewcommand{\footnoterule}{}  

\caption{Equations from the $WBQ$ fit to the different ETGs samples (see Figure 2) and difference between virial and stellar mass. The mean error in minimum, maximum and mean values of the difference between log$({\bf M_{virial}/{\bf M_{\odot}}})$ and log$({\bf M_{*}/{\bf M_{\odot}}})$ is approximately 0.12.}

\scalebox{0.35}{
\begin{tabular}{|l|c|c|c|c|c|}

\hline
\hline
                           &                             &                 &                                         &               &  \\
{\Huge Name of the sample} & {\Huge Equation of the fit} & {\Huge}         &{\Huge Difference between virial and stellar mass} &{\Huge}        &  \\
                           &                             &                 &{\Huge  }         &{\Huge}        &  \\
\hline
\hline
                           &                             &                 &                                         &               &  \\
                           &  {\Huge Total samples}      & {\Huge Minimum} &{\Huge Maximum}                          & {\Huge Mean}  &  \\
                           &                             &                 &                                         &               &  \\
\hline
                           &                             &                 &                                         &               &  \\
 {\Huge de Vaucouleurs Salpeter-IMF stellar mass} & {\Huge $log({\bf M_{*}}/{\bf M_{\odot}}) =  (0.885 \pm 0.085)\; log({\bf M_{virial}}/{\bf M_{\odot}}) + (0.905 \pm 0.093)$ }  &{\Huge 0.187} &{\Huge 0.532} & {\Huge 0.360} &{\Huge a} \\

{\Huge S\'ersic Salpeter-IMF stellar mass} & {\Huge $log({\bf M_{*}}/{\bf M_{\odot}}) =  (0. 861\pm 0.086)\; log({\bf M_{virial}}/{\bf M_{\odot}}) + (1.140 \pm 0.133)$  }  &{\Huge 0.180} &{\Huge 0.597} & {\Huge 0.389} &{\Huge b}\\

{\Huge Kroupa-IMF stellar mass} & {\Huge $log({\bf M_{*}}/{\bf M_{\odot}}) =  (0.874 \pm 0.083)\; log({\bf M_{virial}}/{\bf M_{\odot}}) + (1.101 \pm 0.084)$  }  &{\Huge 0.096} &{\Huge 0.474} & {\Huge 0.285} &{\Huge c}\\
                           &                             &                 &                                         &               &  \\

\hline
                           &                             &                 &                                         &               &  \\

  & {\Huge Morphological samples}  &  &  &  &   \\
                             &                             &                 &                                         &               &  \\
\hline
                           &                             &                 &                                         &               &  \\
 {\Huge de Vaucouleurs Salpeter-IMF stellar mass} & {\Huge $log({\bf M_{*}}/{\bf M_{\odot}}) =  (0.912 \pm 0.086)\; log({\bf M_{virial}}/{\bf M_{\odot}}) + (0.577 \pm 0.100)$  }  &{\Huge 0.259} &{\Huge 0.523} & {\Huge 0.391} &{\Huge d}\\

{\Huge S\'ersic Salpeter-IMF stellar mass} & {\Huge $log({\bf M_{*}}/{\bf M_{\odot}}) =  (0.858 \pm 0.091)\; log({\bf M_{virial}}/{\bf M_{\odot}}) + (1.184 \pm 0.192)$  } &{\Huge 0.167} &{\Huge 0.594}  & {\Huge 0.380} &{\Huge e}\\

{\Huge Kroupa-IMF stellar mass} & {\Huge $log({\bf M_{*}}/{\bf M_{\odot}}) =  (0.902 \pm 0.084)\; log({\bf M_{virial}}/{\bf M_{\odot}}) + (0.794 \pm 0.089)$  }  &{\Huge 0.137} &{\Huge 0.431} & {\Huge 0.284} &{\Huge f}\\
                           &                             &                 &                                         &               &  \\
\hline
                           &                             &                 &                                         &               &  \\

  & {\Huge Homogeneous samples}  &  &  &  &   \\
                             &                             &                 &                                         &               &  \\
\hline
                           &                             &                 &                                         &               &  \\
 {\Huge de Vaucouleurs Salpeter-IMF stellar mass} & {\Huge $log({\bf M_{*}}/{\bf M_{\odot}}) =  (0.815 \pm 0.086)\; log({\bf M_{virial}}/{\bf M_{\odot}}) + (1.658 \pm 0.100)$  }  &{\Huge 0.0995} &{\Huge 0.654} & {\Huge 0.377 } &{\Huge g}\\

{\Huge S\'ersic Salpeter-IMF stellar mass} & {\Huge $log({\bf M_{*}}/{\bf M_{\odot}}) =  (0.845 \pm 0.088)\; log({\bf M_{virial}}/{\bf M_{\odot}}) + (1.324 \pm 0.154)$  }  &{\Huge 0.150} &{\Huge 0.616} & {\Huge 0.383} &{\Huge h}\\

{\Huge Kroupa-IMF stellar mass} & {\Huge $log({\bf M_{*}}/{\bf M_{\odot}}) =  (0.847 \pm 0.085)\; log({\bf M_{virial}}/{\bf M_{\odot}}) + (1.410 \pm 0.095)$  }  &{\Huge 0.0435} &{\Huge 0.5025} & {\Huge 0.273} &{\Huge i}\\
                           &                             &                 &                                         &               &  \\

\hline

\end{tabular}}
\end{table*}


\begin{figure*}

\begin{center}

 \includegraphics[width=17cm]{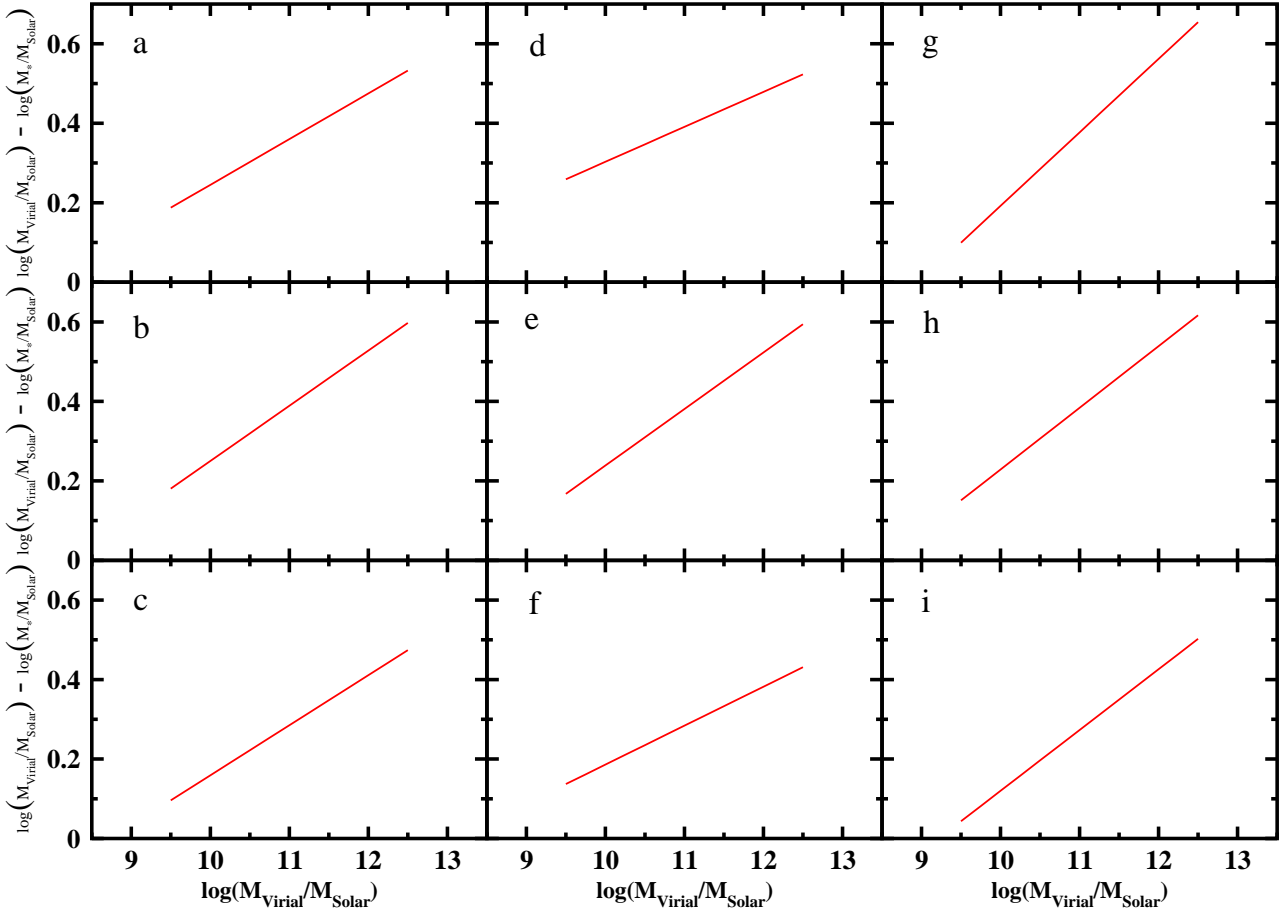}

     \caption{Difference between virial and stellar mass as function of virial mass for the ETGs samples. The first row corresponds to the de Vaucouleurs Salpeter-IMF stellar mass, the second one corresponds to the S\'ersic Salpeter-IMF stellar mass and the third one corresponds to the Kroupa-IMF stellar mass. The mean error in the difference between log$({\bf M_{virial}/{\bf M_{\odot}}})$ and log$({\bf M_{*}/{\bf M_{\odot}}})$ is approximately 0.12.}
         \label{Fig3}

  \end{center}

 \end{figure*}


\begin{figure*}

\begin{center}

  \includegraphics[angle=00,width=17cm]{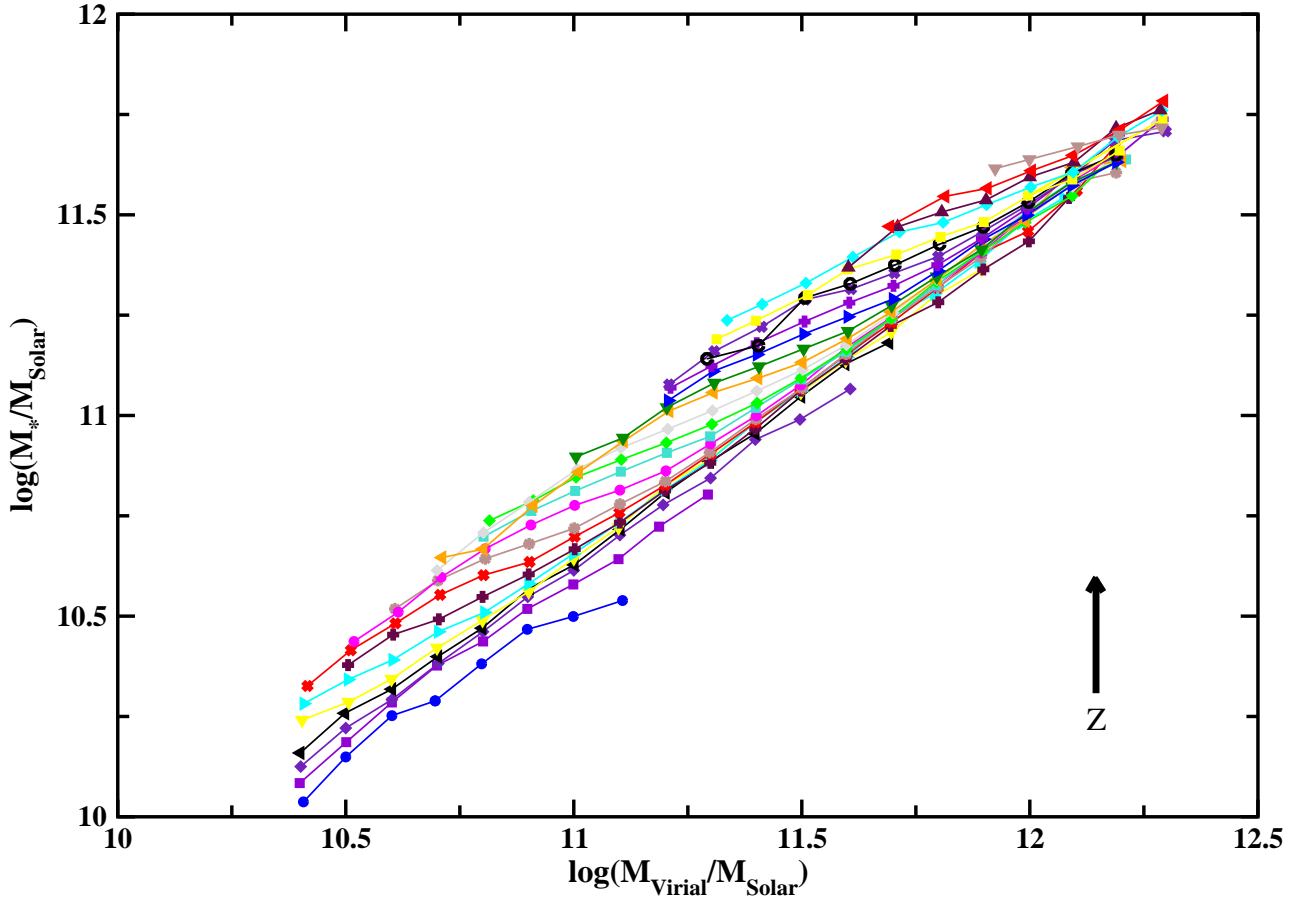}

     \caption{Behaviour of stellar mass as function of virial mass for constant redshift. Each colour and symbol represents a constant redshift. The lower-left part of the graph (blue dots) corresponds to z $\sim$ 0.025 while the upper-right part of the graph (brown triangles) corresponds to z $\sim$ 0.26. The difference in redshift between consecutive symbols is approximately 0.01. The mean errors for log$({\bf M_{virial}/{\bf M_{\odot}}})$ and log$({\bf M_{*}/{\bf M_{\odot}}})$ are approximately 0.052 and 0.065 respectively. }
         \label{Fig4}

  \end{center}

 \end{figure*}


\begin{figure*}

\begin{center}

  \includegraphics[angle=00,width=17cm]{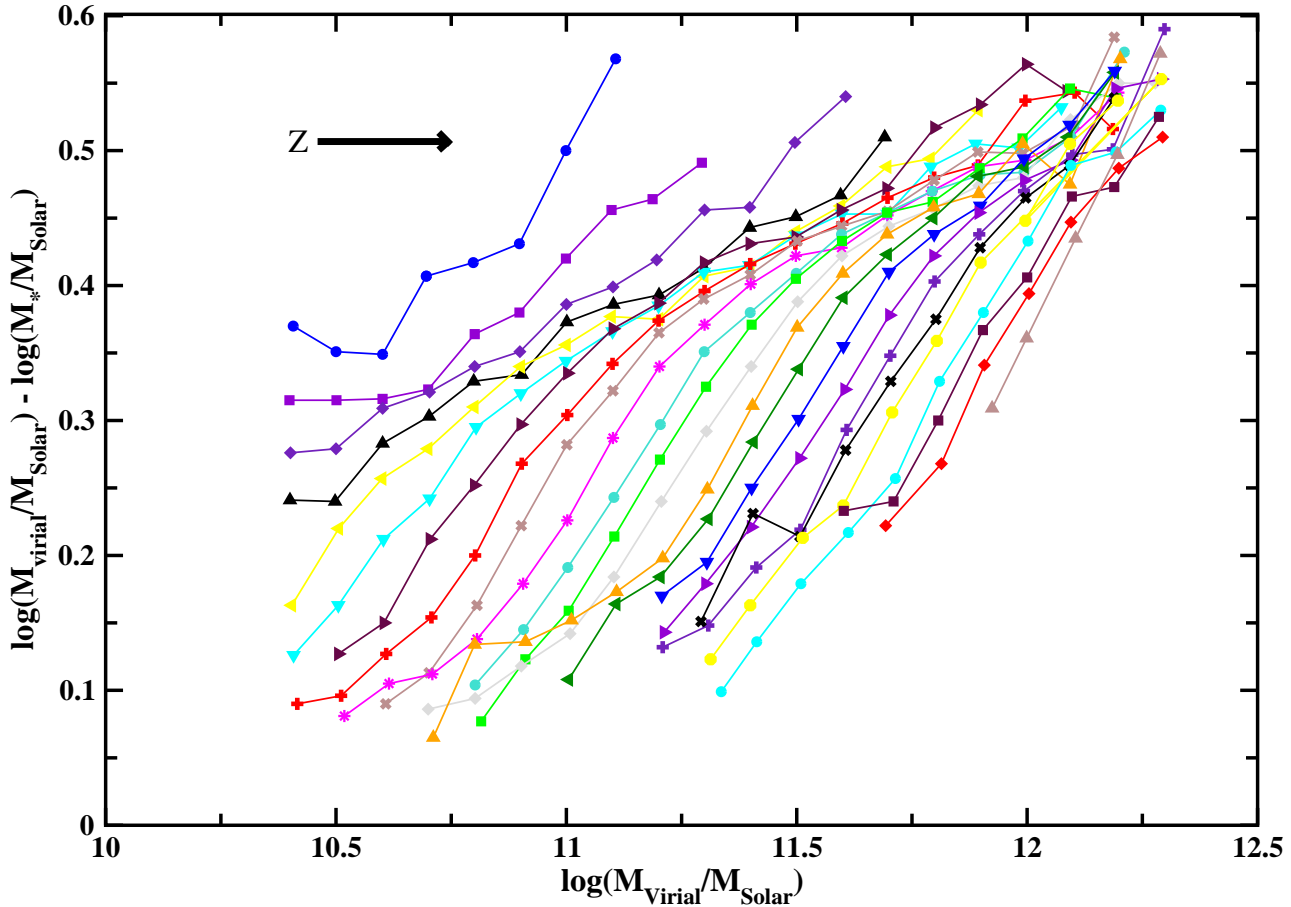}

     \caption{Difference between virial and stellar mass as function of virial mass for the ETGs samples. Each colour and symbol represents a constant redshift. The upper-left part of the graph (blue dots) corresponds to z $\sim$ 0.025 while the lower-right part of the graph (brown triangles) corresponds to z $\sim$ 0.26. The difference in redshift between consecutive symbols is approximately 0.01. The mean error of the difference between log$({\bf M_{virial}/{\bf M_{\odot}}})$ and log$({\bf M_{*}/{\bf M_{\odot}}})$ is approximately 0.12.}
         \label{Fig5}

  \end{center}

 \end{figure*}


\begin{figure*}

\begin{center}

  \includegraphics[angle=00,width=17cm]{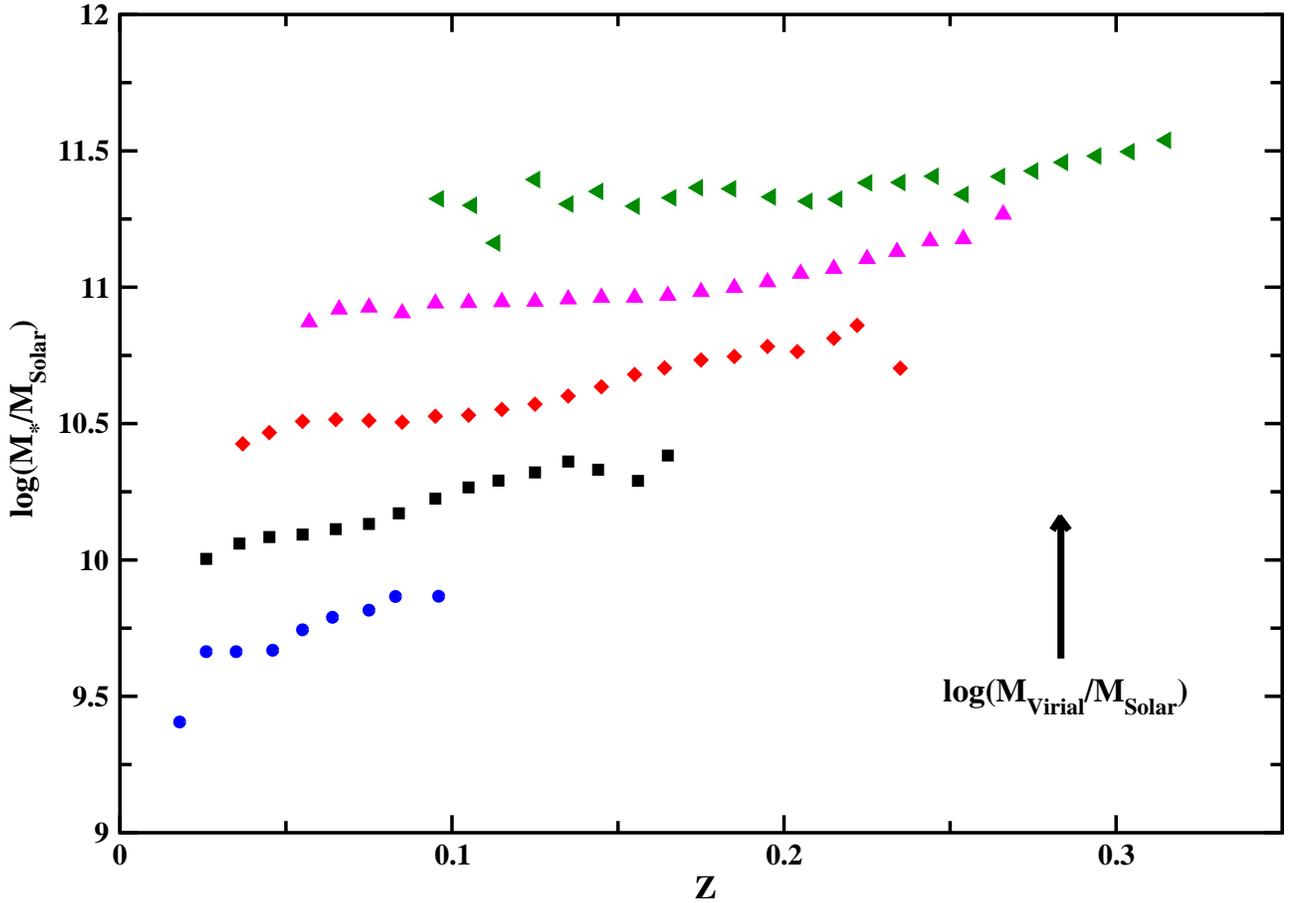}

     \caption{Behaviour of stellar mass as function of redshift for constant virial mass. Each colour and symbol represents constant virial mass. The lower-left part of the graph (blue dots) corresponds to log$({\bf M_{virial}/{\bf M_{\odot}}})$ $\sim$ 10 while the upper-right part of the graph (green triangles) corresponds to log$({\bf M_{virial}/{\bf M_{\odot}}})$ $\sim$ 12. The difference in log$({\bf M_{virial}/{\bf M_{\odot}}})$ between consecutive symbols is approximately 0.5. The mean error of the log$({\bf M_{*}/{\bf M_{\odot}}})$ is approximately 0.065.}
         \label{Fig6}

  \end{center}

 \end{figure*}

\section{Differences between virial and stellar mass. Procedure 1}

In the previous section we have mentioned that there are different procedures to analyse the properties of our ETGs samples. In particular, we can study the difference between the virial and stellar mass using the mass distribution of the samples and the $WBQ$ fit. To obtain the parameters of the linear regression ($WBQ$ fit) for the different ETGs samples we use the Nigoche-Netro et al. (2015) procedure as follows:

\begin{itemize}

\item We calculate the mean value of the logarithm of stellar mass at quasi-constant logarithm of virial mass.

\item We calculate the mean value of the logarithm of virial mass at quasi-constant logarithm of stellar mass.

\item We perform a linear regression ($WBQ$ fit) to the previously mentioned mean values.

\end{itemize}

The term quasi-constant mass in this context means mass intervals in the logarithm of width equal to 0.1.

In Figure 2 we show a mosaic of the behaviour of the stellar mass with respect to the virial mass for the ETGs Total (Column 1), Morphologic (Column 2) and Homogeneous (Column 3) samples. The first row corresponds to the de Vaucouleurs Salpeter-IMF stellar mass, the second one corresponds to the S\'ersic Salpeter-IMF stellar mass and the third one corresponds to the Kroupa-IMF stellar mass. Each graph shows the mean values of the luminous mass distribution at quasi-constant virial mass (black dots), the mean values of the virial mass distribution at quasi-constant stellar mass (blue squares) and the $WBQ$ fit (dashed line) to both point distributions. The solid line is the one-to-one line.

In Table 1 we show the results of the $WBQ$ fit to the different samples of ETGs. The difference between virial and stellar mass shown has been obtained considering the slope and the zero point of the fit for each sample. For each value of virial mass we calculated the stellar mass and the average, maximum and minimum differences among these masses. The mean error in the average, maximum and minimum values of the difference between log$({\bf M_{virial}/{\bf M_{\odot}}})$ and log$({\bf M_{*}/{\bf M_{\odot}}})$ is approximately 0.12. From Table 1 we find the following:

\begin{itemize}

\item	The average difference between log$({\bf M_{virial}/{\bf M_{\odot}}})$ and log$({\bf M_{*}/{\bf M_{\odot}}})$ considering the samples with Salpeter-IMF stellar mass is 0.380, whereas if we consider Kroupa-IMF stellar mass profiles the average difference is 0.281. This seems to indicate that the estimated difference between masses is affected by the IMF used in the calculation of luminous mass. However the mentioned difference seems to be due to the zero point because the slopes of the samples are similar.

\item	Considering only those samples where the masses were obtained using de Vaucouleurs profiles the average difference is 0.376, whereas if we consider S\'ersic profiles the average difference is 0.384. This seems to indicate that the estimated difference between masses is not affected by the profile used in the calculation of luminous and virial mass. This result is confirmed by the slopes of the samples which are similar.
	
\item	If we consider only samples in large intervals of redshift (total and morphological samples) the average difference is 0.348, whereas if we consider only the sample restricted in redshift (homogeneous sample) the average difference is 0.344. This seems to indicate that the average difference between masses is similar when we move from a wide to a narrow redshift interval. However the slopes of the samples seems to refute this result.

\end{itemize}

The previous results are in agreement with the results found by \citet{nig15} . However, the comparison of the mean values of the difference in masses could mask the real behaviour of the samples because the midpoints of two straight lines could be similar even if the slopes of those straight lines are different. So it is necessary to compare the masses in a different way considering that the difference in mass could depend on other variables such as mass and/or redshift.

From Table 1 we can see that, if we consider all the samples, the average difference between the maximum and minimum values for
log(${\bf M_{virial}/{\bf M_{\odot}}})$ - log$({\bf M_{*}/{\bf M_{\odot}}}$) is 0.401. This relatively large difference as well as the equations corresponding to Table 1 suggest that there is a dependence of the log(${\bf M_{virial}/{\bf M_{\odot}}})$ - log$({\bf M_{*}/{\bf M_{\odot}}}$) on virial and stellar mass. This behaviour can be easily seen in Figure 3 where, for all samples, the difference between virial and stellar mass depends on virial mass. From Figure 3, it is interesting to note that the behaviour of the samples with different IMF-stellar mass is similar, although with different zero points. A similar behaviour also occurs when we consider different profiles. Also from Figure 3, it is important to note that the slope for the restricted in redshift sample (homogeneous sample) is steeper than the slope for the samples in large intervals of redshift. That is to say, the difference between virial and stellar mass seems to depend on redshift. The behaviour of the differences between virial and stellar mass as a function of mass and redshift requires a deeper analysis which we will address in the following section.

\section{Differences between virial and stellar mass. Procedure 2}

In the previous section we have found that the difference between virial and stellar mass depends on mass and seems to depend on redshift. To investigate these dependences in a deeper way we can analyse the difference in masses considering quasi-constant mass (virial and stellar) and quasi-constant redshift. The term quasi-constant mass means mass interval in the logarithm of width equal to 0.1. The term quasi-constant redshift means redshift interval equal to 0.01.

In Figure 4 we can see the behaviour of the stellar mass as a function of virial mass considering quasi-constant redshift for the Kroupa-IMF stellar mass sample. Each colour and symbol represents quasi-constant redshift. The redshift value goes from approximately zero (lower-left part of the graph) to approximately 0.3 (upper-right part of the graph). In this figure we can see that for the same value of the virial mass, the stellar mass grows with redshift, that is to say, the difference between virial and stellar mass diminishes with redshfit. We can also see that for the high-redshift and high-mass regime the dispersion of the distribution is lower than in the low-mass regime of the samples. From this figure we can see that the redshift plays an important role in the intrinsic dispersion seen in Figure 1.

In Figure 5 we show the difference between virial and stellar mass as a function of virial mass considering quasi-constant redshift for the Kroupa-IMF stellar mass sample. The redshift value goes from approximately zero (upper-left part of the graph) to approximately 0.3 (lower-right part of the graph). In this graph we can see that for the same value of the virial mass the difference between virial and stellar mass diminishes with redshift. We can also see that the more massive galaxies have a greater difference between virial and stellar mass and that for the high-redshfit and high-mass regime the dispersion of the distribution is lower than in the low-mass regime of the samples. The difference between virial and stellar mass at different redshift is directly related with the dispersion seen in Figures 1 and 4.

In Figure 6 we show the behaviour of the stellar mass as function of redshift considering quasi-constant virial mass for the Kroupa-IMF stellar mass sample. The logarithm of virial mass value goes from approximately 10 (lower-left part of the graph) to approximately 12 (upper-right part of the graph). In this graph we can see that for a constant value of virial mass the stellar mass increases as function of redshift. We can also see that for the high-mass regime the behaviour of the stellar mass as a function of redshift is less steep than in the low-mass regime of the samples. The slope of the stellar mass as function of redshift at quasi-constant virial mass is related with the intrinsic dispersion seen in Figures 1 and 4.

From the previous results we can conclude that the redshift plays an important role in the intrinsic dispersion of the distribution of log(${\bf M_{virial}/{\bf M_{\odot}}})$ vs. log$({\bf M_{*}/{\bf M_{\odot}}}$). We also can conclude that the difference between virial and stellar mass, in the redshift range $0.0024 < z < 0.3500$ and in the dynamical mass range $9.5 < log({\bf M_{virial}}/{\bf M_{\odot}}) < 12.5$, depends on mass and redshift. The difference between dynamical and stellar mass increases as a function of dynamical mass and decreases as a function of redshift. This last result is in agreement with recent works from the literature where it is shown that the amount of dark matter could depend on mass \citep{tor12,cap13} and redshift \citep{tor14}.

If we convert the data shown in figure 5 to percentages, we find that the difference between masses goes from almost zero to approximately 70\% of the virial mass. This difference could be due to the dark matter and/or a non universal IMF. Therefore, the amount of dark matter, in the redshift range $0.0024 < z < 0.3500$ and in the dynamical mass range $9.5 < log({\bf M_{virial}}/{\bf M_{\odot}}) < 12.5$, goes from almost zero to 70\% of the virial mass depending on mass and redshift and on the impact of the IMF on the stellar mass estimation.

It is important to note that we have found similar results for the de Vaucouleurs Salpeter-IMF sample and S\'ersic Salpeter-IMF Sample.

\section{The relation of the Differences between virial and stellar mass with the Fundamental Plane}

The results previously described can be analysed in the Fundamental Plane (FP) context. During the last 30 years a lot of scientific papers about the FP have been published \citep{djo87,dre87}. The FP is a relation among the variables; effective radii ($\log {\kern 1pt} (r_{e} )$), the effective mean surface brightness ($<\mu>_{e}$) and the central velocity dispersion (log $\sigma _{0}$), as follows:

\begin{equation}
\log\,(r_{e})\, \; =\, \; a\,\log\,(\sigma_{0})\, \; +\; b\, \, <\mu>_{e}\;
+\; c,
\end{equation}

where $a$, $b$ and $c$ represent scale factors.

This relation seems to be due to the virial equilibrium of ETGs, however the theoretical and observational results do not agree. The difference between the theoretical and observational results is known as the tilt of the FP. There are different explanations for the FP tilt, for example, it could be due to: the increase of M/L with L \citep{dre87,cio96}, the variation of the FP parameters with redshift \citep{vando96,wuy04}, the variation in the homology of ETGs (non-constant K in the virial relation -see equation 1-) \citep{don13,per15}, the variation of the shape of the light profile and the content and concentration of dark matter relative to luminous matter \citep{cio96} among others. Some of these works have found that the tilt could be due to a combination of several of the mentioned effects \citep{cio96,pru97,pah98,don13} which seems to be the most plausible explanation.

Given that the FP relates dynamic variables and stellar formation processes as log(${\bf M_{virial}/{\bf M_{\odot}}})$ vs. log$({\bf M_{*}/{\bf M_{\odot}}}$) does, we can extrapolate our findings to the FP. In this sense, the mass and redshift dependence of log(${\bf M_{virial}/{\bf M_{\odot}}})$ - log$({\bf M_{*}/{\bf M_{\odot}}}$) found in this work go in the same direction as the increase of the content of dark matter relative to luminous matter along FP and the variation of the FP parameters with redshift. That is to say, our findings go in the same direction of the `hybrid' explanation to the tilt of the FP. However we have to take into account that in this work we consider that the dark matter follows the light and that the scale factor K in the virial relation (see equation 1) depends only on the light profile which, according to some authors \citep{cio96,koopmans2006,thomas2011} is not necessarily appropriate. In a forthcoming paper we will analyse these variables and their relation with the difference between log(${\bf M_{virial}/{\bf M_{\odot}}})$ and log$({\bf M_{*}/{\bf M_{\odot}}}$).

\section{Conclusions}

The analysis of the distribution of stellar mass with respect to virial mass on several samples of ETGs from the SDSS DR9 in the redshift range $0.0024 < z < 0.3500$ and in the dynamical mass range $9.5 < log({\bf M_{virial}}/{\bf M_{\odot}}) < 12.5$ has yielded the following results:

\begin{itemize}

	\item A significant part of the intrinsic dispersion of the distribution of log(${\bf M_{virial}/{\bf M_{\odot}}})$ vs. log$({\bf M_{*}/{\bf M_{\odot}}}$) is due to redshift (see Fig. 4).
	\item The difference between dynamical and stellar mass depends on mass and redshift.
	\item The difference between dynamical and stellar mass increases as a function of dynamical mass and decreases as a function of redshift.
	\item The difference between dynamical and stellar mass goes from almost zero to approximately 70\% of the virial mass depending on mass and redshift (see figure 5). This difference is due to dark matter or a non-universal IMF or a combination of both.
	\item The amount of dark matter inside ETGs would be equal to or less than the difference between dynamical and stellar mass depending on the impact of the IMF on the stellar mass estimation.

\end{itemize}

The previous results have been analysed in the FP context and we have found that they go in the same direction as some FP results found in the literature in the sense that they could be interpreted as an increase of dark matter along the FP and a dependence of the FP on redshift. However in this work we have considered that the dark matter follows the same density profile as the stellar component and that the scale factor K in the virial relation (see equation 1) depends only on the light profile which, according to some authors \citep{cio96,koopmans2006,thomas2011}, is not appropriate for massive and compact galaxies respectively. In a forthcoming paper we will analyse these variables and their possible relation with the log(${\bf M_{virial}/{\bf M_{\odot}}})$-log$({\bf M_{*}/{\bf M_{\odot}}}$) difference.

\section*{Acknowledgments}

To the memory of Mrs. Eutiquia Netro Castillo, an extraordinary woman.

We thank and acknowledge the comments made by an anonymous referee, they improved greatly the presentation of this paper. We also thank to Instituto de Astronom\'{\i}a y Meteorolog\'{\i}a (UdG, M\'exico) and Instituto de Astronom\'{\i}a (UNAM, M\'exico) for all the facilities provided for the realisation of this project. A. Nigoche-Netro and G. Ramos-Larios acknowledge support from CONACyT and PRODEP (M\'exico). Patricio Lagos is supported by a PostDoctoral Grant SFRH/BPD/72308/2010, funded by FCT (Portugal) and Funda\c{c}\~{a}o para a Ci\^{e}ncia e a Tecnologia (FCT) under project FCOMP-01-0124-FEDER-029170 (Reference FCT PTDC/FIS-AST/3214/2012), funded by the FEDER program. A. Ruelas-Mayorga thanks Direcci\'on General de Asuntos del Personal Acad\'emico, DGAPA at UNAM for financial support under project number PAPIIT IN103813. A. M. Hidalgo-G\'amez thanks Instituto Polit\'ecnico Nacional SIP20161416 for financial support under project number SIP20161416.

\end{document}